\documentclass[%
 reprint, superscriptaddress, nobibnotes,
 amsmath,amssymb, aps, prl
]{revtex4-2}

\usepackage{graphicx}
\usepackage{bm}
\usepackage{float}
\usepackage{hyperref}
\hypersetup{hidelinks}
\hypersetup{colorlinks = true, linkcolor=blue, urlcolor=blue,citecolor=blue}
\usepackage{physics}
\usepackage{suffix}
\usepackage{xstring}
\usepackage{siunitx}
\newcommand*{\refform}[1]{%
	\IfBeginWith{#1}{fig:}{Fig.~\ref{#1}}{}%
}
\newcommand*{\Rb}[1]{\ensuremath{\mathrm{^{#1}Rb}}}

\def \metre{~\mathrm{m}}

\def \nanoK{~\mathrm{nK}}

\def \cD{\mathcal{D}}
\def \m2D{\mathrm{2D}}
\def \mExp{\mathrm{Exp}}
\def \mMC{\mathrm{MC}}

\begin{document}

\preprint{APS/123-QED}

\title{Observation of the BKT Transition in a 2D Bose Gas via Matter-Wave Interferometry}

\author{S. Sunami}%
\email{shinichi.sunami@physics.ox.ac.uk}
\affiliation{Clarendon Laboratory, University of Oxford, Oxford OX1 3PU, United Kingdom}

\author{V. P. Singh}
\affiliation{Institut f\"ur Theoretische Physik, Leibniz Universit\"at Hannover, Appelstra{\ss}e 2, 30167 Hannover, Germany}
\affiliation{Zentrum f\"ur Optische Quantentechnologien and Institut f\"ur Laserphysik, Universit\"at Hamburg, 22761 Hamburg, Germany}

\author{D. Garrick}%
\affiliation{Clarendon Laboratory, University of Oxford, Oxford OX1 3PU, United Kingdom}

\author{A. Beregi}%
\affiliation{Clarendon Laboratory, University of Oxford, Oxford OX1 3PU, United Kingdom}

\author{A. J. Barker}%
\affiliation{Clarendon Laboratory, University of Oxford, Oxford OX1 3PU, United Kingdom}

\author{\\K. Luksch}%
\affiliation{Clarendon Laboratory, University of Oxford, Oxford OX1 3PU, United Kingdom}

\author{E. Bentine}%
\affiliation{Clarendon Laboratory, University of Oxford, Oxford OX1 3PU, United Kingdom}

\author{L. Mathey}
\affiliation{Zentrum f\"ur Optische Quantentechnologien and Institut f\"ur Laserphysik, Universit\"at Hamburg, 22761 Hamburg, Germany}
\affiliation{The Hamburg Centre for Ultrafast Imaging, Luruper Chaussee 149, Hamburg 22761, Germany}

\author{C. J. Foot}%
\affiliation{Clarendon Laboratory, University of Oxford, Oxford OX1 3PU, United Kingdom}

\date{\today}

\begin{abstract}
	
	We probe local phase fluctuations of trapped two-dimensional (2D) Bose gases using matter-wave interferometry. This enables us to measure the phase correlation function, which changes from an algebraic to an exponential decay when the system crosses the Berezinskii-Kosterlitz-Thouless (BKT) transition.
	We determine the temperature dependence of the BKT exponent $\eta$ and find the critical value $\eta_c = 0.17(3)$ for our trapped system.
	Furthermore, we measure the local vortex density as a function of the local phase-space density, which shows a scale-invariant behaviour across the transition.
	Our experimental investigation is supported by Monte Carlo simulations and provides a comprehensive understanding of the BKT transition in a trapped system. 
\end{abstract}

 \maketitle
	
	One of the most intriguing phase transitions is the Berezinskii-Kosterlitz-Thouless (BKT) transition, which lies within the XY universality class \cite{Berezinskii1972,Kosterlitz1973}.
	Two-dimensional (2D) systems in this universality class display quasi-long-range order at non-zero temperatures below the transition, whereas true long-range order is precluded by thermal fluctuations \cite{Mermin1966,Hohenberg1967}. 
	Above the transition, the system forms a disordered state. 
	This transition is characterized by the first-order correlation function  $g_1(\bm{r},\bm{r'})=\langle \Psi^{\dagger}(\bm{r})\Psi(\bm{r'}) \rangle$, where $\Psi(\bm{r})$ is the bosonic field operator at location $\bm{r}$, which changes from algebraic scaling $\sim r^{-\eta}$ in the superfluid phase, to exponential scaling in the thermal phase, with universal exponent $\eta_{\mathrm{BKT}}=0.25$ at the transition. 
	The origin of this change is the BKT mechanism, which consists of the unbinding of vortex-antivortex pairs at the phase transition, underscoring the topological nature of the transition. The unbound vortices are strong phase defects and suppress the quasi long-range order.
	The BKT transition occurs in a wide range of physical systems such as liquid helium \cite{Bishop1978}, superconducting films \cite{Epstein1981}, Josephson junction arrays \cite{Resnick1981}, ultracold atoms \cite{Hadzibabic2006} and polariton condensates \cite{Caputo2018}.

	Ultracold atoms have enabled detailed studies of the BKT transition. These systems have a wide range of trappable quantum liquids, that have tunable interactions and consist of bosonic or fermionic particles. This has led to the observation of coherence and superfluid properties \cite{Hadzibabic2006,Desbuquois2012,Luick2020,Sobirey2021,Choi2012,Kwon2015,Rossi2016}, universal scaling behaviour \cite{Hung2011}, and thermally activated vortices \cite{Hadzibabic2006,Choi2013}. 
	The BKT transition in a harmonically trapped 2D quantum gas was studied via matter-wave interferometry \cite{Hadzibabic2006} and via measurement of the momentum distribution \cite{Murthy2015,Fletcher2015}. 
	The microscopic description of order in such an inhomogeneous system typically invokes the local density approximation (LDA), which relates the observed phenomenon to the universal description of a uniform system. 
	The LDA is essential for the understanding of scale-invariant and superfluid properties of inhomogeneous 2D systems \cite{Hung2011,Desbuquois2012}.
	Ref.\ \cite{Boettcher2016} suggested that the LDA can also be used to describe the correlation properties of an inhomogeneous 2D gas, which was not applied in previous measurements where only integrated quantities were measured.
	This resulted in saturation of the exponent at low temperatures \cite{Hadzibabic2006} as well as a critical exponent five times larger than that predicted for $\eta_{\mathrm{BKT}}$ \cite{Murthy2015}.
	To disentangle the spatial inhomogeneity of the system and the universality of the BKT transition in ultracold atomic systems, it is crucial to access the local fluctuations of the system, rather than global or integrated observables.

	In this Letter, we report the direct observation of local phase fluctuations by a selective probing of the relative phase between a pair of 2D Bose gases. We use matter-wave homodyning and selectively image a slice of the interference pattern \cite{Barker2020} to access \textit{local} phase fluctuations.
	This critical improvement allows direct measurement of the phase correlation function and the local vortex density, which are essential to characterise the BKT transition, as we demonstrate in this paper.  
	We identify the critical temperature of the BKT transition by a change in the functional form of the correlation function, which results in a measured value of critical exponent $\eta_c =0.17(3)$ for our trapped system.
	We examine the temperature dependence of the thermally created vortices, which proliferates above the critical temperature and exhibits a scale-invariant behavior.
	We have benchmarked these experimental measurements by carrying out Monte-Carlo (MC) simulations.

\begin{figure}[ht]
\includegraphics[width=0.99\linewidth]{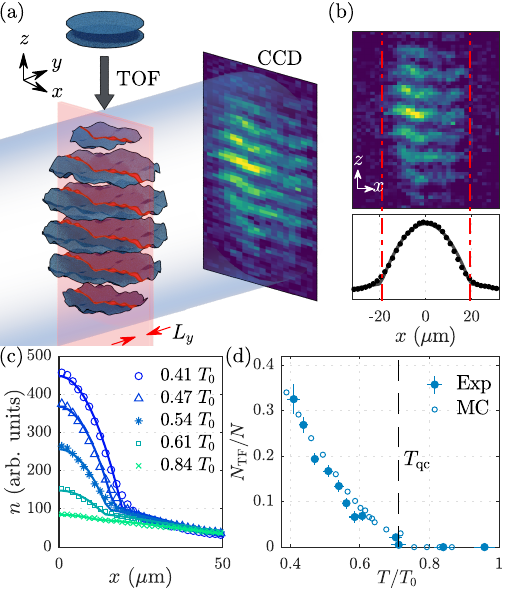}
\caption{\label{fig1} Probing local phase fluctuations using matter-wave interference. 
	(a) Schematic of the experimental procedure. We begin with quasi-2D Bose gases trapped in a double-well potential (blue discs, top). The clouds fall and undergo time-of-flight (TOF) expansion, such that they spatially overlap and produce interference fringes with fluctuating phases (blue wavy planes). 
	The red sheet of thickness $L_y$ denotes the thin laser beam that repumps a slice of atoms.
	We image repumped atoms using resonant light (depicted as a blue beam propagating along the $y$ direction).	
	(b) An example of a single matter-wave interference image (top) and averaged density profile obtained by integrating along $z$ (bottom). The red dash-dotted lines indicate the boundaries of the Thomas-Fermi (TF) region of quasicondensate; see text.
	Grey solid line is the result of bimodal fit.
	(c) Density profiles at different temperatures, where the continuous lines are bimodal fits; see text. 
	(d) Fraction of atoms in the TF profile of the cloud from the experiment (filled markers) and the Monte-Carlo (MC) simulation (open markers).
	The error bars of the experimental results denote standard errors.
}
\end{figure}

\begin{figure}[htt]
	\includegraphics[width=0.99	\linewidth]{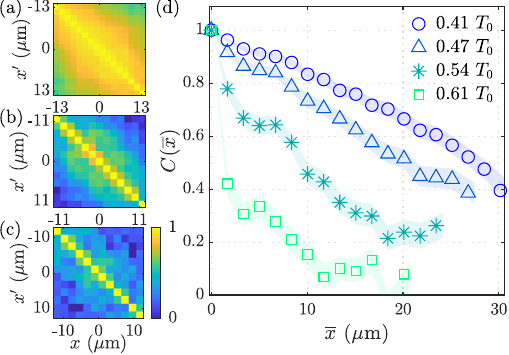}
	\caption{\label{fig:corrfunc} Correlation properties of 2D Bose gases.
		(a-c) Two-point phase correlation function $C^r(x,x')$, obtained from 220 images, at temperatures of $T/T_0=0.41, 0.51$ and 0.59, respectively. 
		(d) Correlation function $C(\overline{x})$ at $T/T_0=0.41, 0.47, 0.54$ and 0.61, from top to bottom. 
		The shaded region corresponds to the standard errors and the range of $\overline{x}$ depends on the temperature because of the change of TF diameter; see text.
	}
\end{figure}

	Our experimental apparatus is described in detail in Refs. \cite{Harte2018,Barker2020}. A cloud of approximately $7 \times 10^4$ \Rb{87} atoms is confined in a cylindrically-symmetric 2D potential, with weak trap frequencies of $\omega_r/2\pi = \SI{11}{\Hz}$ in the horizontal plane and a double-well potential in the vertical direction $z$. 
	The double-well has vertical trap frequencies of $\omega_z/2\pi  = \SI{1}{kHz}$ for each potential minimum and each well confines $N \approx 3.5 \times 10^4$ atoms.
	For the range of temperatures and atom numbers in this work, the quasi-2D conditions $\hbar \omega_z > k_B T$ and $\hbar \omega_z > \mu$ are satisfied, where $\hbar$ is the reduced Planck constant, $k_B$ the Boltzmann constant, $T$ is the temperature of the cloud and $\mu$ is the chemical potential. 
	The characteristic dimensionless 2D interaction strength is $\tilde{g} = \sqrt{8\pi} a_s/\ell_0=0.076$, where $a_s$ is the 3D scattering length and $\ell_0=\sqrt{\hbar/(m\omega_z)}$ is the harmonic oscillator length along $z$ for an atom of mass $m$.
	After loading into the double-well, the gas is held for $\SI{500}{ms}$ to allow equilibration \cite{Hadzibabic2006}. The final temperature of the gas is in the range $\SI{31}  -  \SI{47}{\nano \kelvin}$, which is controlled by forced evaporation. 
	We set the temperature scale for our system using the condensation temperature of an ideal 2D Bose gas in a harmonic trap, $T_0=
	\sqrt{6N}(\hbar \omega_r/\pi k_B) \approx 75 \nanoK$.

	The detection scheme is illustrated in Fig.\ \ref{fig1} (a). 
	To observe the matter-wave interference, the trap is abruptly turned off, releasing the pair of 2D gases for a time-of-flight (TOF) expansion of duration $t_{\text{TOF}}= \SI{16.2}{ms}$.
	Once released, the clouds expand along the $z$ direction \cite{Merloti2013,Hechenblaikner2005} and an interference pattern along $z$ appears.	
	We image a thin slice of the density distribution with thickness $L_y = \SI{5}{\micro\metre}$, as indicated in Fig.\ \ref{fig1} (a).
	%We obtain the density profile by integrating along $z$ and averaging over many images, see Fig.\ \ref{fig1} (b). 
	The density profile, such as that shown in Fig.\ \ref{fig1} (b), is  obtained by integrating along $z$ and averaging over many images. This profile is a bimodal distribution having a Thomas-Fermi (TF) profile of the quasicondensate and a broad Gaussian of the thermal wings \cite{Clade2009}, as illustrated in Fig.\ \ref{fig1} (c).
	In contrast to the true (3D) condensate, the quasicondensate in 2D displays fluctuating phase and supports both thermal and superfluid phases of the BKT transition \cite{Clade2009,Posazhennikova2006,Prokofev2002,Druten1997}.
	We determine the quasicondensate fraction (number of atoms within the TF profile divided by the total number of atoms) and show its temperature dependence in Fig.\ \ref{fig1} (d), along with the MC results \cite{supp}, which yields the onset of quasicondensation at $T_{\mathrm{qc}}/T_0 \sim 0.7$.
	
%	%Once released, the clouds expand along the $z$ direction \cite{Merloti2013,Hechenblaikner2005}, reducing the density quickly so that the interactions are negligible during the expansion. 
%	The radial density distribution after TOF displays a bimodal distribution with a narrow inverted parabola of Thomas-Fermi (TF) peak and a broad Gaussian of the thermal wings \cite{Clade2009}, as shown in Fig.\ \ref{fig1} (b).
%	The appearance of the TF peak is attributed to the significant reduction of the density fluctuation and signals the appearance of quasicondensate \cite{Prokofev2001,Clade2009} where the matter-wave interference was observed.
%	We show the temperature dependence of the density distribution in Fig.\ \ref{fig1} (c). 
%	The fraction of atoms found in the TF peak of the density distribution is plotted in Fig.\ \ref{fig1} (e), along with the Monte Carlo (MC) simulation \cite{supp} result which shows the onset of quasicondensate at $T_{\mathrm{qc}}/T_0 \sim 0.7$.

\begin{figure*}[ht]
	\includegraphics[width=0.99	\textwidth]{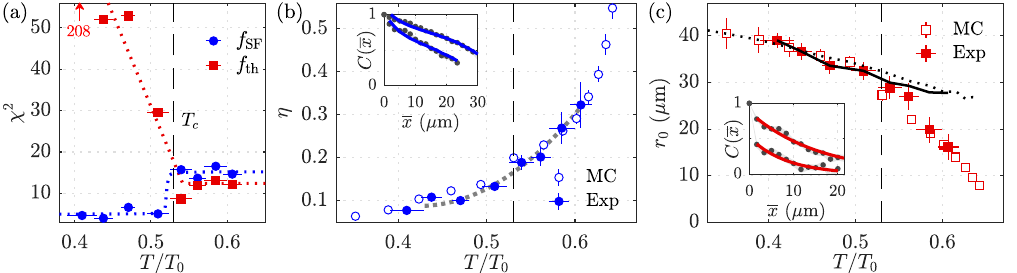}
	\caption{\label{fig:correlation} Characterising the BKT transition in a 2D Bose gas. 
		(a) $\chi^2$ values of the algebraic fit $f_{\mathrm{SF}}$ (blue circles) and exponential fit $f_{\mathrm{th}}$ (red squares) for various values of the temperature $T/T_0$.
		%Smaller values of $\chi^2$ signal a good fit. 
		The dotted lines are the heuristic fits to the temperature dependence of $\chi^2$ to identify the critical point, with arctangent (blue) and a piecewise linear function (red) \cite{supp}.
		The obtained critical temperature $T_c/T_0 = 0.53(1)$ is indicated by the vertical dashed line.
		(b) Measurements of the algebraic exponent $\eta$ (filled circles) are compared with the results of Monte-Carlo simulations (open circles).
		$\eta$ is determined by fitting the correlation function with an algebraic model $f_{\mathrm{SF}}$.
		The grey dotted line is the quadratic fit to the experimental data used to obtain $\eta_c$.
		The inset shows fitted data in the superfluid regime, at $T/T_0 = 0.41$ and 0.51.
		(c) Measurements of the correlation length $r_0$ (filled squares) and the simulation results (open squares), 
		where $r_0$ is determined by fitting the correlation function with an exponential model $f_{\mathrm{th}}$. 
		The values of the temperature-dependent Thomas-Fermi diameter are shown for the experiment (continuous line) and the simulation (dotted line). 
		The error bars in $\eta$ and $r_0$ denote standard fit errors, while the error bars in temperature are statistical errors.
		The inset shows fitted data in the thermal regime, at $T/T_0 = 0.54$ and 0.61.
	}
\end{figure*}

%The wavenumber of the interference fringes along the $z$ direction is consistent with $k = md/\hbar t_{\text{TOF}}$ \cite{Pethick2008}, where $d=\SI{7}{\micro \metre}$ is the spatial separation between the double-well minima.
The local fluctuations of the interference fringes contain the phase information of \textit{in situ} clouds.
At each location $x$, we fit the interference pattern with the function \cite{Pethick2008}
\begin{equation}\label{eq:fringefit}
\rho_x(z) = \rho_0 \exp\left(-z^2/2\sigma^2\right) \left[ 1 + c_0 \cos(kz+\theta(x)) \right],
\end{equation}
where $\rho_0,\sigma, c_0,k,\theta(x)$ are fit parameters. The extracted phase $\theta(x)$ encodes a specific realisation of the fluctuations of the \textit{in situ} local relative phase between the pair of 2D gases. 
For each experimental run, we calculate the two-point phase correlation function $e^{i [\theta(x)-\theta(x')]}$ at locations $x$ and $x'$. 
We then determine the averaged correlation function
\begin{equation} \label{eq:corrmap}
C_{\mathrm{exp}}(x,x') = \frac{1}{N_r} \sum_j e^{i [\theta(x)-\theta(x')]},
\end{equation}
where the index $j$ runs over $N_r$ individual experimental realisations with $N_r=220$. 
We analyse the real part of the correlation function $C^r(x,x')=\text{Re}\left[C_{\mathrm{exp}}(x,x')\right]$, which is equal to $1$ for perfectly correlated pairs of points and $0$ for uncorrelated pairs of points.
Fig.\ \ref{fig:corrfunc} (a-c) show examples of $C^r(x,x')$, which is related to the one-body correlation function $g_1(\bm{r},\bm{r}')$ via \cite{supp}
\begin{equation}
C^r(\bm{r},\bm{r}') \simeq \frac{\langle \Psi^{\dagger}(\bm{r})\Psi(\bm{r'}) \rangle^2}{\langle |\Psi(\bm{r})|^2 \rangle \langle |\Psi(\bm{r'})|^2 \rangle} = \frac{g_1({\bm{r},\bm{r}'})^2}{n_{\text{2D}}^2},
\end{equation}
where $n_{\mathrm{2D}}$ is the 2D density. 
To quantify the decay of correlations, we calculate $C(\overline{x})$ by averaging $C^r(x,x')$ over points with the same spatial separation $\overline{x} = x-x'$ \cite{supp}. 
This averaging was performed over a central region corresponding to 80$\%$ of the TF diameter.
 
The measurements of $C(\overline{x})$ for various temperatures shown in \refform{fig:corrfunc} (d) indicate that $C(\overline{x})$ decays slowly at short and intermediate distances for $T/T_0= 0.41$ and $0.47$ but rapidly at increasing distance at higher temperatures of $T/T_0= 0.54$ and $0.61$. 
This qualitative change of the correlation decay with temperature indicates the crossover to the thermal phase across the BKT transition.
$C(\overline{x})$ falls off rapidly at large distances for all temperatures because of the decrease in density towards the boundary of the TF region.
This effect of density variation can be incorporated into the BKT picture within the LDA by introducing a spatially varying exponent $\eta_t(\overline{x}) = \eta \max(n(\overline{x}))/n(\overline{x})$ \cite{Boettcher2016,supp}, where the density contribution $n(\overline{x}) = \left\langle \sqrt{n(x)n(x+\overline{x})}\right\rangle$ and $\eta$ correspond to the averaged value within the TF region.

To characterize the transition point, we fit the correlation function $C(\overline{x})$ with two models: 
the algebraic model $f_{\mathrm{SF}}(\overline{x}) = a \overline{x}^{- 2\eta_t(\overline{x})}$, where $a$ and $\eta$ are the fit parameters, and the exponential model $f_{\mathrm{th}}(\overline{x})=be^{-2 \overline{x}/r_0}$, where $b$ and $r_0$ are the fit parameters. 
We show the $\chi^2$ values of both fit models in Fig.\ \ref{fig:correlation} (a). 
We observe a transition to exponential scaling at $T_c/T_0=0.53(1)$; 
below $T_c$, $f_{\mathrm{SF}}$ is favored, while above $T_c$, $f_{\mathrm{th}}$ better describes the correlation decay having more than a factor of two higher $p$-values \cite{supp}.

The value of $T_c$ increases when the analysis is limited to narrower regions with higher mean density \cite{supp} and we expect $T_{c,\mathrm{centre}}/T_0=0.68(4)$ in the limit of a small analysis region using the MC results.
This is close to the theoretical prediction of the critical temperature for harmonically-trapped quasi-2D Bose gases within LDA, $T_{c,\mathrm{q2D}}/T_0=0.74$ \cite{Holzmann2008} which is defined as the temperature at which the superfluid appears at the centre of the cloud.
The result of $T_{c,\mathrm{centre}}$ also agrees with the quasicondensation temperature $T_{\mathrm{qc}}/T_0 \sim 0.7$.
% and is close to the quasicondensatation temperature $T_{\mathrm{qc}}\sim 0.7$.
Similarly, at $T_c$, we observe the phase-space density (PSD) $\mathcal{D}= n \lambda^2$ at the trap center $\mathcal{D}_c=$15(2) where $\lambda = h/\sqrt{2\pi m k_B T}$ is the thermal de Broglie wavelength.
In the limit of small regions of interest, we expect $\mathcal{D}_{c,\mathrm{centre}} =9(2)$, which is close to the theoretical prediction $\mathcal{D}_c=\ln(380/\tilde{g})=8.5$ \cite{Prokofev2001}. 
%Furthermore, the observed critical point is lower than the beyond-MF theory prediction \cite{Holzmann2008}. 
%To understand the discrepancy, we emphasize that the critical point observed in this work is the temperature at which a significant fraction of atoms within the quasicondensate become superfluid; 
%this is in contrast to argument in Ref.\ \cite{Holzmann2010}, in which the critical point is defined as the temperature at which the superfluid appear at the centre of the trap.

In \refform{fig:correlation} (b), we show the experimentally determined $\eta$ and the simulation results for various temperatures across the transition, which are in good agreement. 
According to BKT theory, $\eta (T)$ scale approximately linearly in the superfluid phase \cite{Kosterlitz1973,Nelson1977}. 
Indeed, our measurement of $\eta(T)$ follows linear dependence for $T \lesssim T_c$, where the system is in the superfluid regime. 
However, above $T_c$, $\eta(T)$ deviates from the linear behaviour and increases more rapidly.

At $T_c$, the algebraic exponent is $\eta_c \equiv \eta(T_c) = 0.17(3)$, which is below the universal value in the thermodynamic limit $\eta_{\mathrm{BKT}}=0.25$. 
For a finite-size system, the transition manifests itself as a smooth cross-over, for which the critical exponent display a smaller value which scales with the system size as $\eta_c(L) = \eta_{\mathrm{BKT}} / (1+0.5/(\ln(L)+C))$, where $L$ is the linear dimension of the system and C is a nonuniversal constant of order unity \cite{Weber1988}. 
For our trapped system, this expression gives $\eta_c(L) \sim 0.21$ using $L \sim R/\xi \sim 30$, where $R \sim \SI{30}{\micro \metre}$ is the TF diameter and $\xi \sim \SI{1}{\micro \metre}$ is the healing length.	
The value of $\eta_c$ is unaffected by the change in the region of interest \cite{supp}.

In \refform{fig:correlation} (c), we show the correlation length $r_0(T)$ and the temperature-dependent TF diameter. 
Since $r_0$ cannot exceed the system size, the value of $r_0$ is bounded by the TF diameter. 
The experimentally determined values of $r_0$ reach this upper bound for $T \lesssim T_c$ in the superfluid phase. 
When the system crosses $T_c$, $r_0$ becomes smaller than the TF diameter.
We note that up to 15$\%$ systematic error in $r_0$ is expected from the limitations of our imaging system in the thermal regime \cite{supp}.
In \refform{fig:correlation} (c) we present the simulation results for $r_0(T)$ and the TF diameter, which show consistent behaviour in agreement with the measurements. 

\begin{figure}[t]
	\includegraphics[width=0.99	\linewidth]{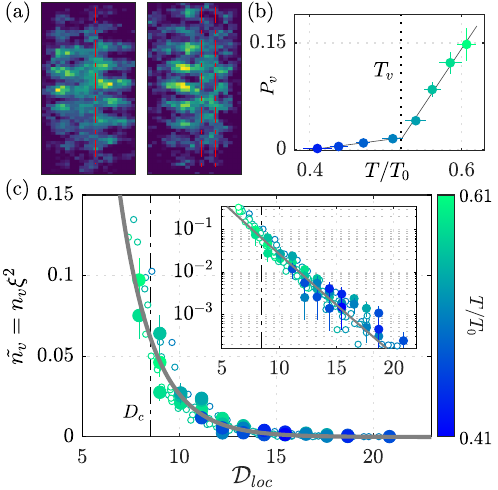}
	\caption{\label{fig:vortex} 
		Vortex proliferation in 2D Bose gases.
		(a) Typical interference patterns with phase dislocations (indicated by red vertical lines), which we count as vortices.
		(b) Probability to detect a vortex $P_v$ as a function of $T/T_0$.  
		The vertical dotted line is the vortex proliferation temperature $T_v$, which is determined using a piecewise linear fit (continuous lines).
		The error bars for $P_v$ are the statistical uncertainty, given by the square root of the numbers of detected vortices.
		(c) The rescaled local vortex density $\tilde{n_v}(x)=n_v(x)\xi(x)^2$ plotted against the local PSD $\cD_{loc}=n(x)\lambda(T)^2$, 
		where $n(x)$ is the local 2D density at location $x$. 
		The measurements (filled circles) and the simulations (open circles) cover a range of temperatures between $T/T_0=0.41$ and $0.61$ and experimental datasets with eight different temperatures contribute to the plot. 
		Solid line is the exponential fit to the experimental data; see text.
		The vertical dash-dotted line is the predicted critical PSD \cite{Prokofev2001}.
		The inset shows the same results on a log-linear scale to highlight the exponential scaling across the BKT transition. Error bars are statistical.
	}
\end{figure}

The BKT transition is driven by thermal vortex unbinding, which suppresses the quasi-long-range order above the critical temperature. 
This underlying mechanism is detected via matter-wave interferometry, where vortices are observed as sharp dislocations in the interference patterns \cite{Hadzibabic2006}. 
This enables us to determine the local vortex density using our selective imaging method \cite{Barker2020,supp}.
In Fig. \ref{fig:vortex} (a), we show examples of matter-wave interference patterns obtained from two independent measurements at $T/T_0=0.52$  and $0.55$. The sharp phase dislocations are indicated by red vertical lines, which we count as vortices. 
We obtain local vortex density $n_v(x)$ by averaging the vortex number over many images at the location $x$ \cite{supp}.
In \refform{fig:vortex} (b), we show the probability to detect a vortex $P_v$, averaged over the TF region; %$P_v$ is small at low temperatures and onset of sharp increase is observed at $T \sim T_c$.
%$P_v$ shows an onset of sharp increase that we determine via a piecewise linear fit. 
%This yields the vortex proliferation temperature $T_v/T_0=$ 0.52(1), in close agreement with $T_c$.
there is a sharp increase in $P_v$ at a certain temperature. We determine this vortex proliferation temperature, $T_v/T_0=$0.52(1), from the discontinuity of the slope in a piecewise linear fit.  

Weakly-interacting 2D Bose gases possess a symmetry which gives rise to the scale-invariant description across the BKT critical point \cite{Pitaevskii1997,Prokofev2001,Hung2011}.
In an inhomogeneous system, local observables can be mapped to a scale-invariant description within LDA, using an appropriate rescaling of the quantities \cite{Hung2011}.
In \refform{fig:vortex} (c) we plot the rescaled local vortex density $\tilde{n_v} = n_v(x)\xi(x)^2$ against the local PSD $\cD_{loc}=n(x)\lambda^2$, where $n(x)$ is the local 2D density at the location $x$ and $\xi(x)=1/\sqrt{n_{\mathrm{qc}}(x)\tilde{g}}$ is the local healing length calculated using the quasicondensate density $n_{\mathrm{qc}}$ \cite{supp}.
The healing length characterises the length scale of a vortex core with its area $\sim \xi^2$, such that $\tilde{n_v}$ quantifies the dimensionless vortex core density.
The measurements for different temperatures collapse on to a common exponential (continuous line), as clearly shown in the inset. 
This is a direct demonstration of scale-invariance of vortex density near the BKT transition. 
The vortex density grows exponentially at low $\cD_{loc}$, indicating crossover to the thermal phase. 
We fit the measured local vortex density with the function $A e^{-\gamma \mathcal{D}_{loc}}$, where $A$ and $\gamma$ are the fit parameters. This choice of exponential scaling is motivated by Ref.\ \cite{Maccari2020}. 
From the fit, we obtain $\gamma_\mExp =0.56(5)$ and $A_{\mExp}=7(2)$.
In \refform{fig:vortex} (c) we also present the simulation result of the vortex density \cite{supp}, which shows scale-invariant behaviour that agrees with the measurements with $\gamma_\mMC=0.52(5)$.
%This agreement is also reflected in the results of the exponential fit, which yields $\gamma_\mMC=0.43(4)$ and $\cD_{c, \mMC}=9(1)$. 
%The values of $\cD_{c, \mExp}$ and $\cD_{c, \mMC}$ are close to the theoretical prediction of the critical PSD $\cD_c = \ln(380/\tilde{g}) = 8.5$ \cite{Prokofev2002}.

In conclusion, we have measured the local phase fluctuations of 2D Bose gases via matter-wave interferometry and supported the measurements by Monte-Carlo simulations.
Our measurements of the phase correlation function and the vortex density provide a comprehensive understanding of the BKT transition in 2D Bose gases.
We identified the critical point by the sudden change in the functional form of correlation function.
We have mapped out the temperature dependence of the algebraic exponent and determined $\eta_c = 0.17(3)$, as expected for a finite-size system.
The local vortex density shows a scale-invariant behavior across the transition. 

The experimental technique presented in this paper can be used to probe non-equilibrium dynamics across the BKT transition \cite{Mathey2010,Mathey2017,Schole2012}. Furthermore, the pair of 2D gases in two trap minima can be coupled via quantum tunneling, to investigate the coupled bilayer XY model \cite{Mathey2007} and Josephson dynamics in low-dimensional quantum gases \cite{Nieuwkerk2019,Zhu2021}.

We acknowledge discussions with Junichi Okamoto, Beilei Zhu and Zoran Hadzibabic. This work was supported by the EPSRC Grant Reference EP/S013105/1. S. S. acknowledges Murata scholarship foundation, Ezoe foundation, Daishin foundation and St Hilda's College for financial support. D. G., A. B., A. J. B. and K. L. thank the EPSRC for doctoral training funding.  
L. M. acknowledges funding by the Deutsche Forschungsgemeinschaft (DFG) in the framework of SFB 925 – project ID 170620586 and the excellence cluster `Advanced Imaging of Matter’ - EXC 2056 - project ID 390715994.
V.P.S. acknowledges funding by the Cluster of Excellence `QuantumFrontiers' - EXC 2123 - project ID 390837967.

 %% supplemental material
 \setcounter{equation}{0}
 \setcounter{figure}{0}
 \setcounter{table}{0}
 \renewcommand{\theequation}{S\arabic{equation}}
 \renewcommand{\thefigure}{S\arabic{figure}}
 
 \newcommand{\figmain}{Fig.\ 1}
 \newcommand{\figcorrfunc}{Fig.\ 2}
 \newcommand{\figeta}{Fig.\ 3}
 \newcommand{\figvortex}{Fig.\ 4}
 
 \newcommand{\eqfringefit}{(1)}
 \newcommand{\eqcorrmap}{(2)}
 \newcommand{\eqlocaleta}{(4)}

 \section*{Supplemental Material}

 \subsection{Monte-Carlo simulation}
 
 \begin{figure*}[t]
 	\includegraphics[width=1\textwidth]{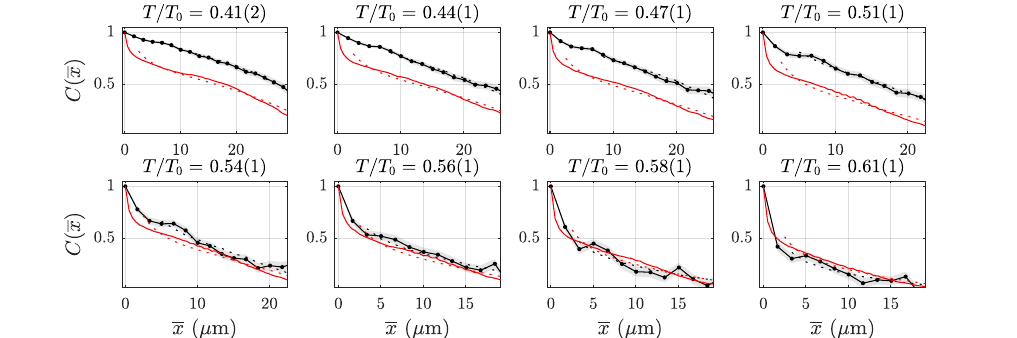}
 	\caption{\label{fig:allcorr} The phase correlation functions $C(\overline{x})$ at eight different temperatures (black connected points) with uncertainties obtained by a bootstrapping method (gray shaded areas). The red line is the correlation function from simulation $C_{\mathrm{sim}}(\overline{x})$ at corresponding temperatures, $T/T_0 = $0.42, 0.46, 0.49, 0.53, 0.53, 0.56, 0.59, 0.6 from left top to right bottom subfigures. Dotted lines with corresponding colours are the fitted power-law model $f_{\mathrm{SF}}$ for each dataset. 
 	}
 \end{figure*}

 We use a classical Monte-Carlo simulation to study the BKT transition in a trapped 2D Bose gas of $^{87}$Rb atoms. 
 The system is described by the many-body Hamiltonian 
 \begin{align} \label{eq_hamil}
 \hat{H} &= \int \mathrm{d}{\bf r} \Big[ \frac{\hbar^2}{2m}  \nabla \hat{\psi}^\dagger({\bf r}) \cdot \nabla \hat{\psi}({\bf r}) + \frac{g}{2} \hat{\psi}^\dagger({\bf r})\hat{\psi}^\dagger({\bf r})\hat{\psi}({\bf r})\hat{\psi}({\bf r})  \nonumber \\
 & \quad + V({\bf r}) \hat{\psi}^\dagger({\bf r})\hat{\psi}({\bf r})   \Big],
 \end{align}
 where $\hat{\psi}$ ($\hat{\psi}^\dagger$) is the bosonic annihilation (creation) operator. The 2D interaction parameter is given by $g = \tilde{g}\hbar^2/m$. 
 We use the value $\tilde{g}=0.076$, which corresponds to the conditions in the experiment. The harmonic trapping potential is $V({\bf r})= m \omega_{r}^2 r^2/2$, where $\omega_{r}$ is the trap frequency and $r=\sqrt{x^2+y^2}$ is the radial coordinate. We use $\omega_{r}=2\pi \times \SI{11}{\hertz}$.  
 
 We simulate this system by mapping it on a lattice system of sites $N_x \times N_y = 200 \times 200$ with a discretisation length of $l=\SI{0.5}{\micro \metre}$. 
 For an accurate representation of the continuum limit, $l$ is chosen to be smaller than, or comparable to, the healing length $\xi = \hbar/\sqrt{2mgn}$ and the thermal de Broglie wavelength (defined in the main text) \cite{Mora2003}.
 In the classical-field approximation, we replace the operators $\hat{\psi}$ by complex numbers $\psi$  \cite{Singh2017}. 
 We then generate the initial states in a grand-canonical ensemble of temperature $T$ and chemical potential $\mu$ 
 via a classical Metropolis algorithm.  
 We choose $\mu$ such that the cloud consists of about $3.5\times 10^4$ atoms, corresponding to the atom number in each well in the experiment. We carry out simulations for a wide range of temperatures to cover the BKT transition.  
 The simulated cloud corresponds to the equilibrium system that is reached in the experiments after a slow splitting and $\SI{500}{\milli \mathrm{s}}$ of equilibration time. 
 For the experimental parameters,  the healing length is in the range $\xi= \SI{0.45} -  \SI{0.61}{\micro \metre}$ and the thermal de Broglie wavelength in the range $\lambda= \SI{0.84} -  \SI{1.1}{\micro \metre}$,  which are larger than or comparable to the used discretization length of $l=\SI{0.5}{\micro \metre}$.
 This ensures that the simulation fulfills the continuum limit, which we have verified in Refs. \cite{Singh2020, Singh2021}. 
 
 For each sample, we calculate the phase $\phi(x)$ and the density $n(x)=|\psi(x)|^2$ using the field $\psi(x)$ along the line $\bm{r}=(x,y=0)$.     
 We then calculate the phase correlation function $C_{\mathrm{sim}}'(\overline{x})$ from phase profiles $\phi(x)$, as described in the main text. We used $N_r^{\mathrm{sim}}=500$ realisations to calculate the $C_{\mathrm{sim}}'(\overline{x})$. To compare with the measurements of the correlation function $C(\overline{x})$, we use $C_{\mathrm{sim}}(\overline{x}) = (C_{\mathrm{sim}}'(\overline{x}))^2$, since fluctuations of two clouds contribute to the relative phase fluctuation that we observe in the experiment. 
 In \refform{fig:allcorr}, we compare $C(\overline{x})$ and $C_{\mathrm{sim}}(\overline{x})$ for various temperatures across the BKT transition. 
 The long-range behavior of the correlation function is similar for both the experiment and the simulation at all temperatures. 
 This results in good agreement of $\eta$ and $r_0$ that we observe in \figeta.
 At low temperatures and short distances, the decay of $C_{\mathrm{sim}}(\overline{x})$ is faster than that of $C(\overline{x})$, which is expected since short-range fluctuations disrupt phases in the simulation, while in the experiment this effect is masked by finite imaging resolution. 
 
 To identify vortices in the simulation we calculate the phase winding around the lattice plaquette of size $l\times l$ using $\sum_{\Box} \delta \phi(x,y) = \delta_x\phi(x,y) + \delta_y\phi(x+l,y)+\delta_x\phi(x+l,y+l)+\delta_y\phi(x,y+l)$, 
 where the index $\Box$ indicates a sum over lattice plaquettes and the phase differences between sites are taken to be $\delta_{x/y} \phi(x,y)  \in (-\pi, \pi]$. 
 We identify a vortex and an antivortex by a phase winding of $2\pi$ and $-2\pi$, respectively. 
 We determine the density profile $n(x)$ and the vortex distribution $n_v(x)$ by averaging them over the region $L_y= \SI{5}{\micro \metre}$ and the ensemble, where $n_v$ counts both a vortex and an antivortex. 
 The resulting values of $n_v (x)$ as a function of the local phase-space density $n(x) \lambda^2$ for various temperatures are shown in \figvortex~(c).

 \subsection{Experimental procedure} \label{sec:expproc}
 
 We form the double-well potential for the dressed atoms using a combination of a static and radiofrequency (RF) magnetic fields \cite{Harte2018}. The static field is a quadrupole magnetic field with cylindrical symmetry about a vertical axis. Three RF fields are applied to give a mutiple-RF (MRF) double-well trap. Control over the amplitudes of RF components allows us to shape the potential from a single well into a double-well potential, as described in Refs. \cite{Harte2018,Bentine2017,Bentine2020,Barker2020jphysb}. In this work, we use RF amplitudes that provide tight confinement in the vertical direction and produce 2D potential. 
 
 After loading the atoms into a single-RF dressed potential and performing evaporative cooling, we transfer the atoms into the MRF-dressed potential. The initial parameters for the MRF-dressed potential are chosen to spatially overlap with the single-RF dressed potential, which minimises heating and atom loss. Subsequently, the RF amplitudes are ramped over $\SI{220}{\milli \second}$ to transform the MRF potential from a single well to a double-well potential, and to increase the confinement along the $z$ direction to $\omega_z/2\pi= \SI{1}{\kHz}$ while radial trapping frequency remains at $\omega_r/2\pi=\SI{11}{\Hz}$, realising a 2D geometry.
 
 After equilibrating the gases for $\SI{500}{\ms}$, the MRF-dressed potential is turned off by first switching off the RF fields and subsequently the quadrupole magnetic field. 
 The switching of RF field projects the internal state of atoms in each well into Zeeman substates labelled by quantum numbers $m_F$. 
 Atoms projected into the $m_F=0$ state are unaffected by the magnetic field and expand ballistically while the other components experience a force from the gradient of the residual magnetic field. 
 Therefore only atoms with $m_F=0$ are important for the analysis of matter-wave interference and temperature measurements \cite{Barker2020}. 
 Populations in $m_F=\pm 1$ states are included in the count when determining the total atom number $N$.
 We choose the phase of the RF at turn-off to ensure equal proportions of atoms are projected to $m_F=0$ from each wells, such that the density profile of atoms in the $m_F=0$ state after TOF gives complete information of the system up to a global rescaling of density. 
 The separation of the wells is $\SI{7}{\micro \metre}$ which is large compared to the characteristic length scale of the cloud along the $z$ direction $\ell_0 \sim \SI{1}{\micro \metre}$ and the two clouds are decoupled. 
 We ensure the populations in the two wells are equal by maximizing the observed matter-wave interference contrast as described in Ref.\ \cite{Barker2020}.  The density scaling factor can be obtained by counting the number of atoms in $m_F=\pm 1 $ components. 
 
 Finally, to locally probe the fluctuating matter-wave interference patterns, we apply a sheet of repumping light before absorption imaging that propagates vertically (in $z$ direction) with thickness $L_y = \SI{5}{\micro \metre}$ and width much larger than the extent of the cloud of atoms \cite{Barker2020}. All atoms are initially in a state with $F=1$, and are then selectively pumped to $F=2$ by the sheet of repumping light, which we image using a light resonant for the atoms in the $F=2$ state. We ensure the repumping light passes through the centre of the cloud by moving the pattern along the $y$ direction, parallel to the propagation of imaging light, to the position where the total absorption signal is maximum.
 
 \subsection{Atom number calibration}
 To calibrate our imaging detectivity, we measure the 3D BEC critical point in a cylindrically symmetric time-averaged adiabatic potential \cite{Gildemeister2012} with trap frequencies $\omega_r/2\pi=\SI{54}{\hertz}$ and  $\omega_z/2\pi=\SI{300}{\hertz}$, resulting in geometric mean $\overline{\omega}/2\pi=(\omega_r \omega_r \omega_z)^{1/3}/2\pi= \SI{96}{\hertz}$. This method is insensitive to experimental imperfection such as imperfect probe light polarisation and provides absolute calibration of atom number. For the range of temperatures and atom numbers used in this process, the system is three-dimensional (not quasi-2D). We calibrate the atom number by comparing the critical atom number to the theoretical value given by
 
 \begin{equation}\label{eq:Nbec}
 N_c = \zeta(3)\left(\frac{k_BT}{\hbar\overline{\omega}}\right)^3\left(\frac{1}{1-3.426(a_s/\lambda)}\right)^3,
 \end{equation}
 where $\zeta$ is the Riemann function. The last term in Eq. \eqref{eq:Nbec} is the mean-field correction \cite{Smith2011}. We have chosen the range of atom numbers used for this calibration such that the average optical density after TOF is close to the values for data shown in the main text.

 \begin{figure*}[ht]
 	\includegraphics[width=0.75\textwidth]{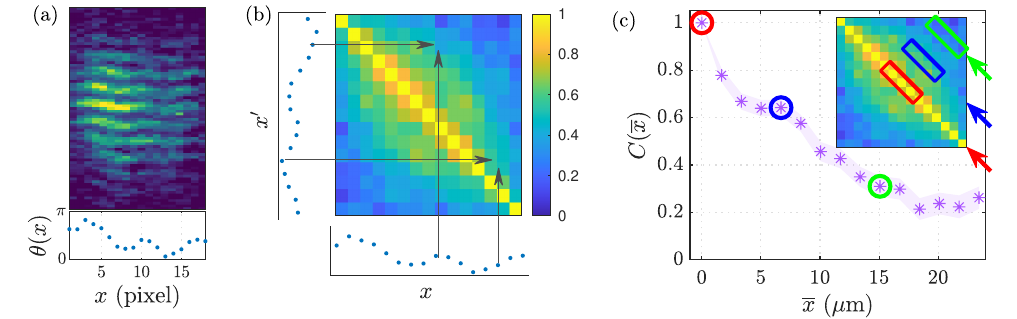}
 	\caption{\label{fig:corranalysis} Extraction of the phase correlation function  $C(\overline{x})$. 
 		(a) An example phase profile extracted from the interference pattern using the fit function Eq. \eqfringefit. The bottom panel shows the phase profile in radians.
 		(b) From the phase profiles $\theta(x)$ extracted from images, we calculate the two-point phase correlation Eq. \eqcorrmap. Diagonal elements $C^r(x,x)$ are always 1 by definition. 
 		(c) From $C^r(x,x')$, the correlation function $C(\overline{x})$ at distance $\overline{x}$ (in number of pixels) is calculated by taking the mean of $\overline{x}$-th diagonal elements of  $C^r(x,x')$ according to Eq. \eqcorrmap. The elements of $C^r(x,x')$ included in the averaging are indicated by the rectangles (red, blue and green), corresponding to the three $C(\overline{x})$ data points marked by circles in matching colors.
 	}
 \end{figure*}

 \subsection{Image analysis procedure}\label{sec:imageanalysis}
 \textit{Preprocessing of images} \ Prior to all image analysis procedures described in this paper, we perform preprocessing of images to minimize the imaging noise, such as optical fringing, using the method described in Ref.~\cite{Ockeloen2010}.\\
 
 \textit{Temperature extraction from density profiles} \ We analyse the temperature of the atom clouds by repeating the experimental runs with all atoms in the cloud (not just a slice) repumped into the detectable hyperfine level, thus recording the integrated density distribution. We fit the density distribution of atoms with $m_F=0$ recorded after 16.2 ms TOF with a bimodal distribution along the $x$ direction. 
 The expansion of the gas along the $z$ direction, which is initially tightly confined in the trap, follows the expansion of a Gaussian wavepacket \cite{Hechenblaikner2005}. In the radial direction $r$, the density distribution is well-described by a bimodal model which is the sum of broad Gaussian and Thomas-Fermi peak (inverted parabola). The Gaussian part expands ballistically during TOF but the Thomas-Fermi peak does not change size significantly over  $t_{\text{TOF}}=\SI{16.2}{ms}$. We extract the temperature of the gas from the fitted value of the Gaussian pedestal after TOF. This procedure also allows us to extract the PSD by estimating the \textit{in situ} 2D density. 
 The mean PSD reported in the main text was obtained by 70$\%$ of peak PSD at each temperatures, which corresponds to the mean PSD within 80$\%$ of 2D TF density profile.
 We repeated experiments at least 14 times for each temperature and fitted individual realisations with the bimodal model. In the main text, the reported temperatures are the mean of fitted temperature values and the error bars for the temperatures denote standard errors.
 
 We note that finite population of excited levels in the transverse harmonic oscillator can affect the thermometry as pointed out in Refs. \cite{Hadzibabic2008,Holzmann2008}. Nevertheless, our experimental parameters are well within the quasi-2D regime as the mean-field and thermal energy scales are smaller than $\hbar \omega_z$ and the population of excited levels is small. The observed agreement of density distributions between experiment and strictly 2D Monte Carlo simulation (see \figmain~(b) and \figeta~(c)) further confirms the results of the thermometry method described above.\\

\begin{figure}[H]
 	\includegraphics[width=0.95\linewidth]{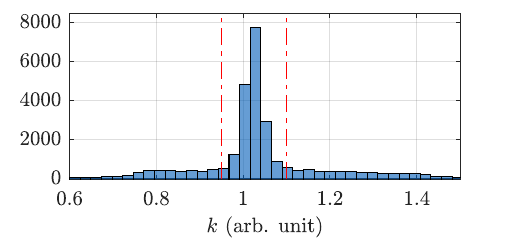}
 	\caption{\label{fig:postselection} Histogram of the fringe wavenumber $k$ from fits of local column densities with Eq. \eqfringefit. The red lines indicate the range of $k$ used for postselection.}
 \end{figure}

 \textit{Extracting the relative phase} \ We analyse the local phase of the matter-wave interference patterns following the procedure outlined in Refs. \cite{Barker2020,Hadzibabic2006}, in which we perform a 1D fit of column density $\rho_x(z)$ at each location $x$ with Eq. \eqfringefit. We discard datapoints for which the fitted value of $k$ lies outside the peak of its histogram (see \refform{fig:postselection}) as this indicates either a lack of signal in the pixel column or failure of the fit procedure. \\ 
 
 \subsection{Correlation analysis}\label{sec:corranalysis}
 
 The spatial correlation of the relative phases $\theta(\bm{r})$ is related to the one-body correlation $g_1(\bm{r})$ in the following way. We consider the two-point correlation of \textit{relative} phases of two independent clouds: 
 \begin{equation}
 C(\bm{r},\bm{r}') := \frac{\langle \Psi_1(\bm{r})\Psi_2^{\dagger}(\bm{r})\Psi_1^{\dagger}(\bm{r}')\Psi_2(\bm{r}')\rangle}{\langle|\Psi_1(\bm{r})|^2\rangle\langle|\Psi_2(\bm{r}')|^2\rangle},
 \end{equation}
 where $\Psi_j(\bm{r})$ are the bosonic field operators at location $\bm{r}$ of clouds ($j=1,2$).	
 We describe $\Psi_j(\bm{r})$ in terms of the density-phase representation as $\Psi_j(\bm{r})=\sqrt{n_j(\bm{r})}e^{i\varphi_j(\bm{r})}$. 
 Assuming uniform and equal density $n_j(\bm{r})\sim n_{\text{2D}}$ for each cloud, as well as small density fluctuations \cite{Hadzibabic2011}, we can simplify the expression $\Psi_j(\bm{r})\sim\sqrt{n_{\text{2D}}}e^{i\varphi_j(\bm{r})}$. 
 Writing the relative phase of the two fields as $\theta(\bm{r})=\varphi_1(\bm{r})-\varphi_2(\bm{r})$,
 \begin{equation}
 C(\bm{r},\bm{r}') = \langle e^{i\theta(\bm{r})-i\theta(\bm{r}'))} \rangle.
 \end{equation}
 If the two clouds in the double-well are decoupled, the fields $\Psi_1(\bm{r})$ and $\Psi_2(\bm{r})$ fluctuate independently. Furthermore, assuming that the two clouds are identical, we find
 \begin{equation}
 C(\bm{r},\bm{r}') \simeq \frac{\langle \Psi^{\dagger}(\bm{r})\Psi(\bm{r'}) \rangle^2}{\langle |\Psi(\bm{r})|^2 \rangle \langle |\Psi(\bm{r'})|^2 \rangle} = \frac{g_1({\bm{r},\bm{r}'})^2}{n_{\text{2D}}^2}.
 \end{equation}

 The averaged form of correlation function $C(\overline{x})$ is defined as
 \begin{equation}\label{eq:corrfunc}
 C(\overline{x}) = \left\langle C^r(x,x+\overline{x}) \right\rangle_{x\in w} ,
 \end{equation}
 where $w = [-\overline{x}/2-2\ell_p,-\overline{x}/2+2\ell_p]$, $\ell_p=\SI{1.67}{\micro \metre}$ is the image-plane pixel size and the analysis is performed within $80\%$ of the Thomas-Fermi diameter.
 \refform{fig:corranalysis} illustrates some further aspects of the correlation analysis method.
 
 \subsection{Critical temperature}
 We identify the critical point by the change in the functional form of the correlation functions.
 We fit the obtained correlation functions $C(\overline{x})$ with algebraic and exponential models $f_{\mathrm{SF}}$ and $f_{\mathrm{th}}$, described in the main text and extract the $\chi^2$-value.
 %Good fits result in smaller values of $\chi^2$, thus providing a quantitative method to compare the fit models.
 At low temperature ($T\lesssim 0.53 T_0$), the $\chi^2$ statistic for the algebraic model is $\sim$5 whereas that for the exponential model is above 25.
 These values and the degree of freedom for the fit gives the corresponding $p$-values (the probability of observing the $\chi^2$ values or higher) of around 0.9 for the algebraic model and below 0.05 for the exponential model:
 the algebraic model is a valid description of the data at temperatures $T\lesssim 0.53T_0$ \cite{Hughes2010}.
 On the other hand, at higher temperatures, the $\chi^2$ values for the algebraic model is $\sim$16 whereas that for exponential model is $\sim$12.
 The $p$ values are around 0.2 and 0.5, respectively;
 while the algebraic model is not rejected (for example at 5$\%$ level), exponential model is a significantly better description of the observed data.
 To obtain the temperature where the sudden change in $\chi^2$ is observed, we performed heuristic fits of the temperature dependence of $\chi^2$ as shown in \figeta (a); for the algebraic model we used an arctangent function 
 \begin{equation}
 f(T,a,b,T_{c1}) = a\arctan[10^3\times(T-T_{c1})]+b.
 \end{equation}
 and for the exponential model a piecewise function defined by
 \begin{equation}
 f(T,a,b,T_{c2}) = 
 \Big\{
 \begin{array}{lr}
 a(T_{c2}-T) + b, & \  \ T \leq T_{c2},\\
 b, & \ \ T > T_{c2}.
 \end{array}
 \end{equation}
 The fit results are shown in \figeta (a).
 From the obtained $T_{c1}$ and $T_{c2}$, we estimate the critical temperature 
 \begin{equation}
 T_c = \frac{T_{c1}+T_{c2}}{2}.
 \end{equation}
 We define the uncertainty of the critical temperature as $\delta T_c = |T_{c1}-T_{c2}|/2$.
 
 The critical algebraic exponent $\eta_c$ was obtained from the fit of the temperature dependence of $\eta$ in \figeta (b) using $f(T)=a+bT+cT^2$.
 We obtain the uncertainty of the value of $\eta_c$ from $\delta \eta_c = |\eta(T_{c1})-\eta(T_{c2})|/2$.

 \subsection{Finite-size effect}
\begin{figure}[hh]
 	\includegraphics[width=0.95\linewidth]{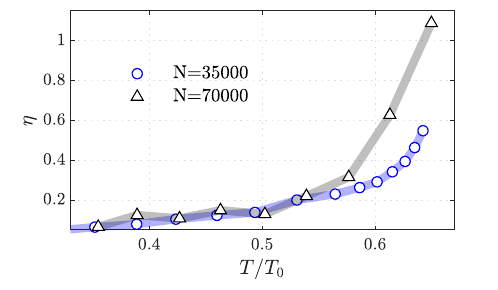}
 	\caption{\label{fig:finitesize} 
 		MC simulation results of $\eta$ at different atom number $N=35000$ (blue, circle) and $N = 70000$ (black, triangle). Lines that connect the points are guide to the eye.
 	}
 \end{figure}

 To demonstrate the finite-size effect on the BKT transition in harmonically-trapped 2D Bose gases, we have performed the Monte Carlo simulation with a larger system size having $N=70000$ atoms, a factor of two larger than the results presented in the main text in which we kept the ideal-gas condensation temperature the same by using $\omega_r/2\pi=9.2$ Hz. 
 This is motivated by the scaling for thermodynamic limit in 2D harmonic trap, $N\rightarrow \infty, \omega_r \rightarrow 0, N\omega_r^2=$const. \cite{Posazhennikova2006}. 
 The crossover behavior near the transition temperature is affected by the finite size of the system, as shown in Fig.\ \ref{fig:finitesize}, and results in sharper increase of the exponent at higher temperature for larger system.
 
 \subsection{Effect of the size of the analysis region}
 The correlation analysis is performed over a finite region of inhomogeneous density profile, in which the local PSD varies. 
 This makes a direct comparison to the theoretical prediction of critical PSD \cite{Prokofev2001} $\mathcal{D}_c = 8.5$ difficult.
 Furthermore, in the theoretical work on the harmonically-trapped 2D gases \cite{Holzmann2008,Holzmann2010}, critical point was defined as the temperature at which the center of the cloud becomes superfluid;
 this is in contrast to the critical temperature obtained in this work, which requires a significant fraction of the system in the analysis region to be in the superfluid regime.
 
\begin{figure}[hh]
 	\includegraphics[width=0.95\linewidth]{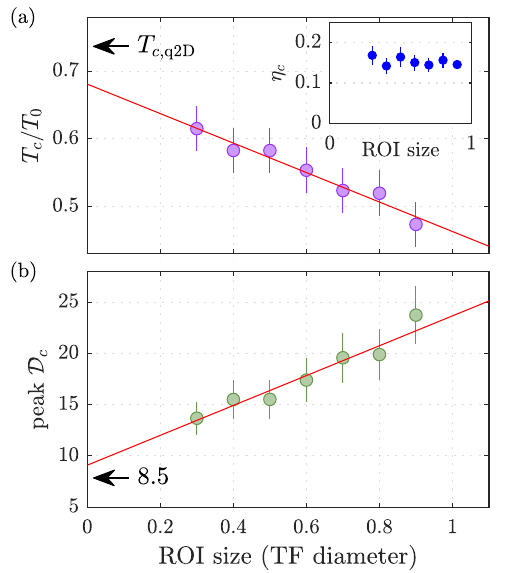}
 	\caption{\label{fig:ROI} 
 		The dependence of the critical point on the region of interest (ROI) for the correlation analysis.
 		(a) the dependence of critical temperature on the ROI size in the unit of TF diameter ($\sim \SI{30}{\micro \metre}$ around the critical point). 
 		Red line is the linear fit to the data, whose intercept gives $T_c = 0.68(4)$.
 		Inset shows the critical algebraic exponent $\eta_c$ which is insensitive to the size of the ROI.% (in the same unit of $x$ axis as the main figure).
 		(b) Peak PSD at the critical temperature obtained from the MC simulation. 
 		The intercept of linear fit (red) is 9(2).
 	}
 \end{figure}

 To demonstrate the effect of these discrepancies and to enable a quantitative comparison, we repeat the correlation analysis of the numerical simulation data within regions of varying size (defined as a fraction of TF region) and the results for critical temperature and the peak PSD at the critical temperature are shown in Fig.\ \ref{fig:ROI}. 
 We find a linear dependence of the values on the size of the region of interest;
 using linear extrapolation of these values, we find the critical point at $T_{c,\mathrm{center}}/T_0 = 0.68(4)$ with peak PSD $\mathcal{D}_{c,\mathrm{center}} = 9(2)$.
 These values are close to the theoretical predictions $T_{c,\mathrm{q2D}}=0.74 T_0$ \cite{Holzmann2008} and $\mathcal{D}_c=8.5$ \cite{Prokofev2001}.
 As such, we may understand the critical point reported in the main text $T_c = 0.53(1)T_0$ as the lower bound of critical point in Refs.\ \cite{Holzmann2008,Holzmann2010}.
 The critical algebraic exponent $\eta_c$ is independent of the analysis region, since this value is dominantly affected by the finite-size effect.
 
 \subsection{Vortex detection} \label{sec:vortexdetection}
\begin{figure}[h]
 	\includegraphics[width=0.95\linewidth]{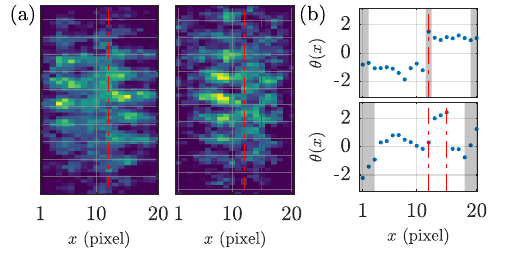}
 	\caption{\label{fig:vortexdetection}  Vortex detection. (a) Example images with vortices.  The image on the left was taken at a temperature of $T/T_0=0.52$ and that on the right at $T/T_0=0.55$. The disruption of the relative phases at different $x$ is evident. (b) Phase profiles extracted from the two images in (a): the top panel is from image on the left and the bottom panel is from the image on the right. Gray shaded regions indicate either outside of Thomas-Fermi region or failed fit at the pixel column (top panel at $x=12$). The vortex detected in the top panel coincides with a grey shaded area; this suggests the presence of vortex core where the phase cannot be defined.
 	}
 \end{figure}

 From the extracted phase profiles, we detect vortices with the same methodology as Ref.\ \cite{Hadzibabic2006} but with the following improvements. 
 Firstly, since our imaging resolution is comparable to the expected size of a vortex core $ \sim \SI{1}{\micro \metre}$ and, furthermore, a single pixel corresponds to an equivalent size, the vortex detection is performed by evaluating the phase difference at second adjacent pixels. 
 Secondly, to avoid the local phases returned from failed fitting being counted as a vortex, we have only evaluated where the postselection criteria, as shown in Fig. \ref{fig:postselection}, is satisfied for the pair of datapoints being evaluated. 
 \refform{fig:vortexdetection} illustrates the vortex detection. The vortex detection was performed within $90\%$ of the TF diameter to probe a wide range of local PSD. 
 
 To obtain local vortex density $n_v(x)$, we first calculate the local vortex probability $p_v(x)$ which is defined by the ratio of the number of detected vortices to the number of phase difference evaluations at location $x$ (datapoints that are rejected by the postselection criteria above are not counted towards the number of phase evaluations).
 The local vortex density $n_v(x)$ can be obtained by dividing $p_v (x)$ by the vortex detection area of a single pixel column, $\ell_p L_y =  \SI{8.4}{\micro \metre}^{2}$ where $\ell_p= \SI{1.67}{\micro \metre}$ is the image-plane pixel size. 
 We further divide the obtained vortex density by a factor of two, due to the fact that the detection of a phase jump corresponds to the existence of a vortex in one of the two layers of the 2D clouds.
 
 To obtain \figvortex~(c), we calculated the \textit{in situ} 2D density of atoms $n(x)$ at location $x$ and used these values for the calculation of the local PSD while the results for healing length was calculated using the Thomas-Fermi peak of the density distribution to infer the local quasicondensate density.
 The vortex data was binned by local PSD for clarity and there are eight experimental datasets with temperatures $T/T_0 = $ 0.41, 0.44, 0.47, 0.51, 0.54, 0.56, 0.59 and 0.62 contributing to the plot.
 %For the inset (log-linear plot), we excluded the datapoints with less than one detected vortex. 
 The errors of the exponential fit results reported in the main text are obtained by a bootstrapping method.

 \subsection{Local correlation approximation} \label{sec:algfittrap}
 
 In harmonically trapped 2D Bose gases, the spatially varying density of the gas modulates local thermodynamic quantities. This feature was utilised to show scale invariance of the system in Ref. \cite{Hung2011}. Such effects introduce deviation from the BKT picture derived in uniform systems. Recently, a spin-wave theory for trapped 2D systems was derived, focusing on the trap effect on the correlation function \cite{Boettcher2016};
 using \textit{local correlation approximation} (LCA) the exponent $\eta$ was expected to vary according to the local density $n(\bm{r})$ as
 \begin{equation}
 g_1(\bm{r},\bm{r}') \propto |\bm{r}-\bm{r}'|^{-\eta \frac{n_0}{\sqrt{n(\bm{r})n(\bm{r}')}}},
 \end{equation}
 where $n_0$ is the peak density. It was found that the LCA reproduces the result of trapped spin-wave theory well \cite{Boettcher2016}. We use this procedure with the minor modification described below to derive $f_{\mathrm{SF}}(\overline{x})$, which we used for the results presented in the main text.
 
\begin{figure}
 	\includegraphics[width=0.95\linewidth]{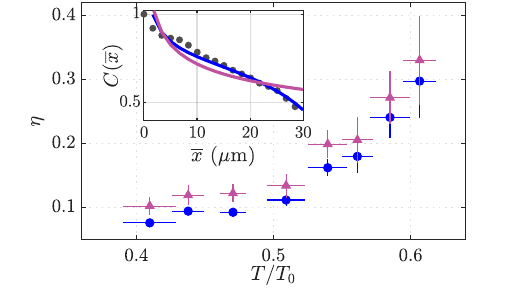}
 	\caption{\label{fig:trapeffect} Effect of inhomogeneity on the correlation function. 
 		The purple diamonds are from fits with the bare algebraic function and the blue circles are from fits using the model function incorporating the inhomogeneity as described in the text. Error bars in temperatures are statistical and error bars in $\eta$ are standard fit errors.
 		(inset) The measured correlation function (black dots) is fitted with $f_{\mathrm{alg}}$ (purple line) and $f_{\mathrm{SF}}$ (blue line) described in the text.
 	}
 \end{figure}

 Since the data points in the correlation functions deduced from the experimental measurements are averaged over multiple locations, we replace $n(r)$ with $n(\overline{x})$ defined by 
 \begin{equation}	
 n(\overline{x}) = \left\langle \sqrt{n(x)n(x+\overline{x})} \right\rangle_{x\in w},
 \end{equation}
 where $w$ is defined in the same way as for the calculation of correlation function in Eq.\ (\ref{eq:corrfunc}). 
 We then use the model function $f_{\mathrm{SF}}(\overline{x}) = a \overline{x}^{-2\eta_t(\overline{x})}$, with $\eta_t(\overline{x}) = \eta\frac{\max(n(\overline{x}))}{n(\overline{x})}$ where $a$ and $\eta$ are fit parameters. 
 For $n(x)$, we used estimated $in \ situ$ density distributions. 
 As shown in the inset of \refform{fig:trapeffect}, the model describes the long-range behavior of the measured correlation decay very well in the superfluid regime. 
 The fitted values of $\eta$ are compared to the fit result using a bare algebraic function, $f_{\mathrm{alg}}=a\overline{x}^{-2\eta}$ in Fig. \ref{fig:trapeffect}. 
 Overall, the scaling of $\eta$ is similar while there is a clear downward shift of the values $\eta$.
 
 \subsection{Effect of finite imaging resolution}

 \textit{Superfluid regime} \
 The imaging system in our apparatus has finite imaging resolution, with a point-spread function (PSF) approximated by a Gaussian with standard deviation $\sigma_{\text{PSF}}=\SI{2.1}{\micro \metre}$. 
 The observed images are thus a density distribution convolved with the PSF of the imaging system. 
 \begin{figure}[t]
 	\includegraphics[width=0.95\linewidth]{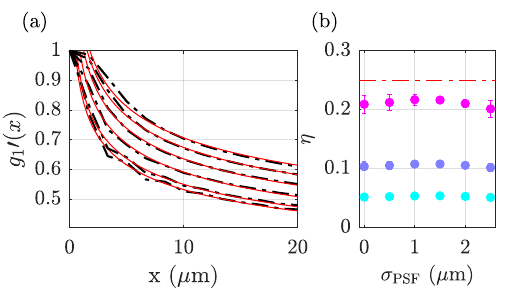}
 	\caption{\label{fig:gaussianblur} Effect of imaging resolution on the measurement of $\eta$.
 		(a) Normalised one-body correlation functions $g_1'(x)=g_1(x)/n$ in the superfluid regime, calculated with the effect of imaging in real-space with $\sigma_{\text{PSF}}$ = 0, 0.5, 1, 1.5, 2, 2.5 $\SI{}{\micro \metre}$ from bottom to top (Black dash-dotted lines). $\sigma_{\text{PSF}}$ = 0 corresponds to the absence of any imaging effect. Red lines are corresponding fit results with $f(x)=a x^{-\eta}$. 
 		(b) Fitted values of $\eta$ as a function of imaging resolution with $\eta_0=0.2$ (magenta), 0.1 (purple) and 0.05 (cyan). The error bars are the 95\% confidence intervals of the fit.
 	}
 \end{figure}
 To model the effect of imaging resolution on the correlation analysis within the superfluid regime, we assume the equilibrium phonon mode population \cite{Hadzibabic2011} and the 
 effect of imaging resolution can be straightforwardly modeled by multiplying the phonon mode populations by $\exp(-\sigma_{\text{PSF}}^2 k^2/2)$. 
 We plot the normalised correlation function with true exponent $\eta=0.2$ in Fig.\ \ref{fig:gaussianblur} (a), along with fitting with power-law decay $f(x)=a x^{-\eta}$. 
 In Fig.\ \ref{fig:gaussianblur} (b) we present the fitted values of $\eta$ which show consistent results over the range of $\sigma_{\text{PSF}}$ and find that finite imaging resolution of $\sigma_{\text{PSF}}=\SI{2.1}{\micro \metre}$ has negligible effect on the extraction of $\eta$ within the superfluid regime.

\begin{figure}[h]
 	\includegraphics[width=0.95\linewidth]{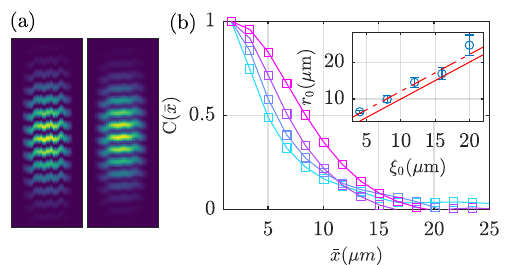}
 	\caption{\label{fig:gaussianblurthermal} The effect of imaging resolution in the thermal regime. 
 		(a) Simulated density distributions after TOF showing matter-wave interference. 
 		The correlation length used to generate the fluctuating phase profile was $\xi_0= \SI{10}{\micro \metre}$. 
 		The left image is the simulated distribution and the right image is the result after a filtering procedure that models the experimental imaging process including finite imaging resolution and pixel size.
 		(b) Phase correlation functions obtained from simulated images, with imaging resolution $\sigma_{\mathrm{PSF}} =$0,1,2,3 $\SI{}{\micro \metre}$ (cyan to magenta) and $\xi_0=\SI{8}{\micro \metre}$.
 		(Inset) Fitted correlation length in the presence of finite imaging resolution $\sigma_{\mathrm{PSF}} = \SI{2.1}{\micro \metre}$, at various values of true correlation length $\xi_0$ for the generation of the fluctuating phase profiles. Error bars are 95$\%$ confidence interval of the fit. 
 		The red solid line is $r_0=\xi_0$ and the red dash-dotted line is $r_0=\xi_0 +\sigma_{\mathrm{PSF}} =\xi_0+\SI{2.1}{\micro \metre}$. 
 	}
 \end{figure}

 \textit{Thermal regime} \
 To estimate the effect of finite imaging resolution on the determination of the correlation function in the thermal regime, we have simulated the fluctuations in the system using the Ornstein-Uhlenbeck stochastic process \cite{Gardiner2009} with the time axis replaced by the real-space axis along a direction perpendicular to the imaging direction. 
 This process gives fluctuating phase profiles with exponentially decaying two-point phase correlation at given correlation length $\xi_0$, and has been used to simulate the fluctuating phase profiles of 1D gases \cite{Stimming2010,Langen2013}. 
 The effect of finite imaging resolution was incorporated by generating the interference pattern assuming the expansion of a pair of phase-fluctuating clouds along the $z$ direction \cite{Pethick2008} and convolving the resulting density distribution with a Gaussian PSF. 
 We generate $N=200$ simulated images for each set of parameters and apply the same fitting procedure as described in the main text to obtain the correlation functions. 
 In \refform{fig:gaussianblurthermal} (b) the correlation functions in the presence of finite imaging effect are plotted. 
 The inset shows the correlation length $r_0$ extracted by fitting the correlation functions in the presence of an imaging effect with $\sigma_{\mathrm{PSF}}=\SI{2.1}{\micro\metre}$ against $\xi_0$, the true correlation length from which the simulated data was generated. We find that the imaging effect systematically shifts the observed correlation length up to $\sigma_{\mathrm{PSF}}$, which corresponds to up to 15$\%$ systematic error for the measured correlation lengths presented in \figeta~(c).

\end{document}